\documentclass[11pt]{article}
\usepackage{blois,epsfig}

\def\be{\begin{equation}}
\def\ee{\end{equation}}
\def\bea{\begin{eqnarray}}
\def\eea{\end{eqnarray}}

 \newcommand{\fr}[2]{\frac{{\displaystyle #1}}{{\displaystyle #2}}}

\begin{document}
\vspace*{4cm}
\title{CONSTRAINING THE DARK 2HDM}

\author{ M. KRAWCZYK and D. SOKO\L OWSKA}

\address{Institute of Theoretical Physics, University of Warsaw,\\
 ul. Ho\.za  69, 00-681 Warsaw, Poland }

\maketitle\abstracts{
The Dark 2HDM (called also the Inert Doublet Model) is the version of the Two-Higgs-Doublet Model
with an exact $Z_2$ symmetry.
It contains the SM-like Higgs boson $h$ and dark scalars $H$,$H^+,H^-$,$A$.
Only dark scalars have odd $Z_2$ parity and therefore the lightest of them  can be a dark matter candidate.
The comparison of such model and
of the standard 2HDM with an explicit $Z_2$ symmetry is given.
}

\section{Introduction: 2HDMs}
There are various versions of Two-Higgs-Doublet Model (2HDM) with two SU(2) scalar doublets $\phi_{1,2}$ leading to very distinct physical phenomena.
One observe a tight relation between symmetry under the  $Z_2$ transformation (change of sign of one doublet) and CP conservation  in the multi-Higgs doublet models.
If $Z_2$ is explicitly conserved in the Lagrangian of 2HDM, then CP is conserved in the 2HDM \cite{branco}. If $Z_2$ is (at least) softly violated ,
then CP can be violated  explicitly or spontaneously.
For a hard $Z_2$ violation  new phenomena,  like FCNC
and CP violation without CP mixing \cite{sok},  may appear   at the tree level.
In addition there are various assignments of the Yukawa interaction with fermion fields (Model I, Model II as realized in MSSM, Model III etc.).

Model with an exact $Z_2$ symmetry, which is conserved both explicitly and spontaneously,  is called  the Inert Doublet Model (IDM) or Dark 2HDM  \cite{ma,bar}.
Here  $\phi_1$ and all known SM  fields are $Z_2$-even, while  $\phi_2$ is $Z_2$-odd and its  vacuum expectation values (vev) have to be equal to zero.
The  first doublet in IDM plays a role identical to the scalar doublet in the SM, being responsible for a generation of masses of gauge bosons and fermions. Here the only Higgs particle is a SM-like Higgs boson $h$, with tree-level couplings to gauge bosons and fermions equal to the corresponding couplings for the SM-Higgs boson. The second scalar doublet has nothing to do with mass generation, nor it has  direct couplings to fermions (vev = 0) -  it is "inert" from this point of view.
Physical particles are scalars  $H, A, {H^+}, {H^-}$ with $Z_2$-odd quantum number.
Since $Z_2$ symmetry is strictly conserved, these particles can  be produced and annihilated only in pairs. The lightest dark scalar is stable
being a candidate for dark matter particle. To be a good WIMP dark matter candidate it should be neutral.

The phenomenology of  the IDM, a valuable model for today \cite{bar,cao,Lundstrom:2008ai,Gustafsson:2007pc,LopezHonorez:2006gr,su}, is very distinct from all other 2HDM
versions, although formally it is similar to the  Model I and  in some aspects it is very close to the SM. Some of the constraints can be derived from the analysis
performed at LEP. Constraints on this model may also come  from the astrophysical data.
\section{The potential and its extrema}\label{subsec:pot}
ƒThe most general potential for two SU(2) doublets with weak hypercharge Y=+1 is given by
\begin{equation}\begin{array}{c}
V=\fr{\lambda_1}{2}(\phi_1^\dagger\phi_1)^2
+\fr{\lambda_2}{2}(\phi_2^\dagger\phi_2)^2+
\lambda_3(\phi_1^\dagger\phi_1) (\phi_2^\dagger\phi_2)\\[4mm]
+\lambda_4(\phi_1^\dagger\phi_2) (\phi_2^\dagger\phi_1)
+\fr{1}{2}\left[\lambda_5(\phi_1^\dagger\phi_2)^2
+h.c.\right]+\Delta V_m^4+{\cal M}(\phi_i)+V_0\,;\\[4mm] \Delta
V_m^4 =\left\{\left[\lambda_6(\phi_1^\dagger\phi_1)+\lambda_7
(\phi_2^\dagger\phi_2)\right](\phi_1^\dagger\phi_2)
+h.c.\right\},\\[4mm]
{\cal M}(\phi_i)=-\fr{1}{2}\left\{m_{11}^2(\phi_1^\dagger\phi_1)+
\left[m_{12}^2 (\phi_1^\dagger\phi_2) +h.c.\right]+
m_{22}^2(\phi_2^\dagger\phi_2)\right\},\quad V_0=const\,.
\end{array}\label{baspot}
\end{equation}
Here $\lambda_{1-4}$, $m_{11}^2$ and $m_{22}^2$ are real (by
hermiticity of potential) while parameters $\lambda_{5-7}$ and
$m_{12}^2$ are generally complex. Taking into account the reparametrization freedom
 one can show that only eleven of them are independent \cite{mk}.
Here we will focus on such form of the potential, where there is no
$(\phi_1,\,\phi_2)$ mixing,
so both $m_{12}^2=0$ and  $\Delta V_m^4=0$.
This can be ensured by imposing on V the discrete $Z_2$ symmetry
under the  transformation:
\be
\phi_1\,\to \phi_1 \,\,, \phi_2\,\to - \phi_2.
\label{eq:z2}
\ee
(The same effect can be obtained by imposing the $Z'$ symmetry:
$\phi_1\,\to -\phi_1 \,\,, \phi_2\,\to \phi_2$.)

By using a freedom of reparametrization we can fix $\lambda_5$ to be real.
Useful  abbreviations are:
 \bea
 \lambda_{345}=\lambda_3+\lambda_4+\lambda_5,\quad
 \tilde{\lambda}_{345}=\lambda_3+\lambda_4-\lambda_5,
\\[2mm]
 \Lambda_{345\pm}=\sqrt{\lambda_1\lambda_2}\pm\lambda_{345},\;\;
 \tilde{\Lambda}_{345\pm}=\sqrt{\lambda_1\lambda_2}\pm\tilde{\lambda}_{345}, \;\; \Lambda_{3\pm}= \sqrt{\lambda_1\lambda_2}\pm\lambda_3.
 \eea

To have a {\it stable vacuum} the potential must be positive at
large quasi--classical values of fields $|\varphi_i|$ ({\sl
{positivity constraints}}) for an arbitrary direction in the
$(\varphi_1,\varphi_2)$ plane \cite{fer}. This condition limits possible values
of $\lambda_i$
in the space ${\cal S}_\lambda$ of parameters $\lambda_i$:
 \bea
{\cal S}_\lambda:\;\; \lambda_1>0\,, \quad \lambda_2>0,\;\,
\Lambda_{3+}>0,\;\,\Lambda_{345+}>0,\;\,\tilde{\Lambda}_{345+}>0.
\label{eq:abbr}
\eea
The extremum fulfilling this positivity condition which has the lowest energy is a global minimum - a true vacuum \cite{kan}.

\begin{figure}
\psfig{figure=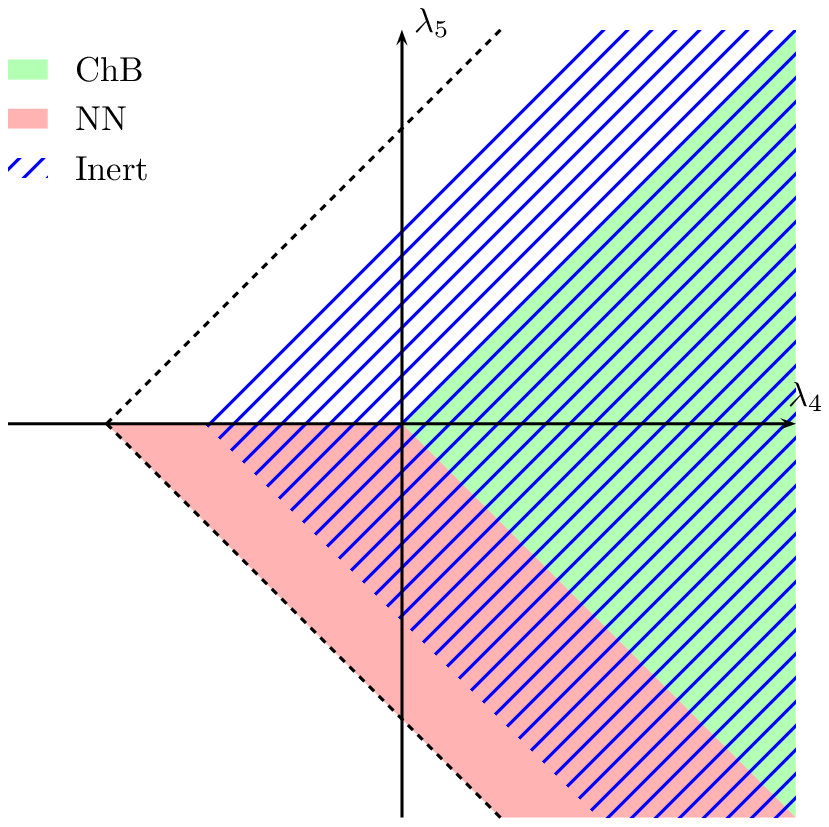,height=7.in}
\vskip -19.cm
\hskip 7cm
\psfig{figure=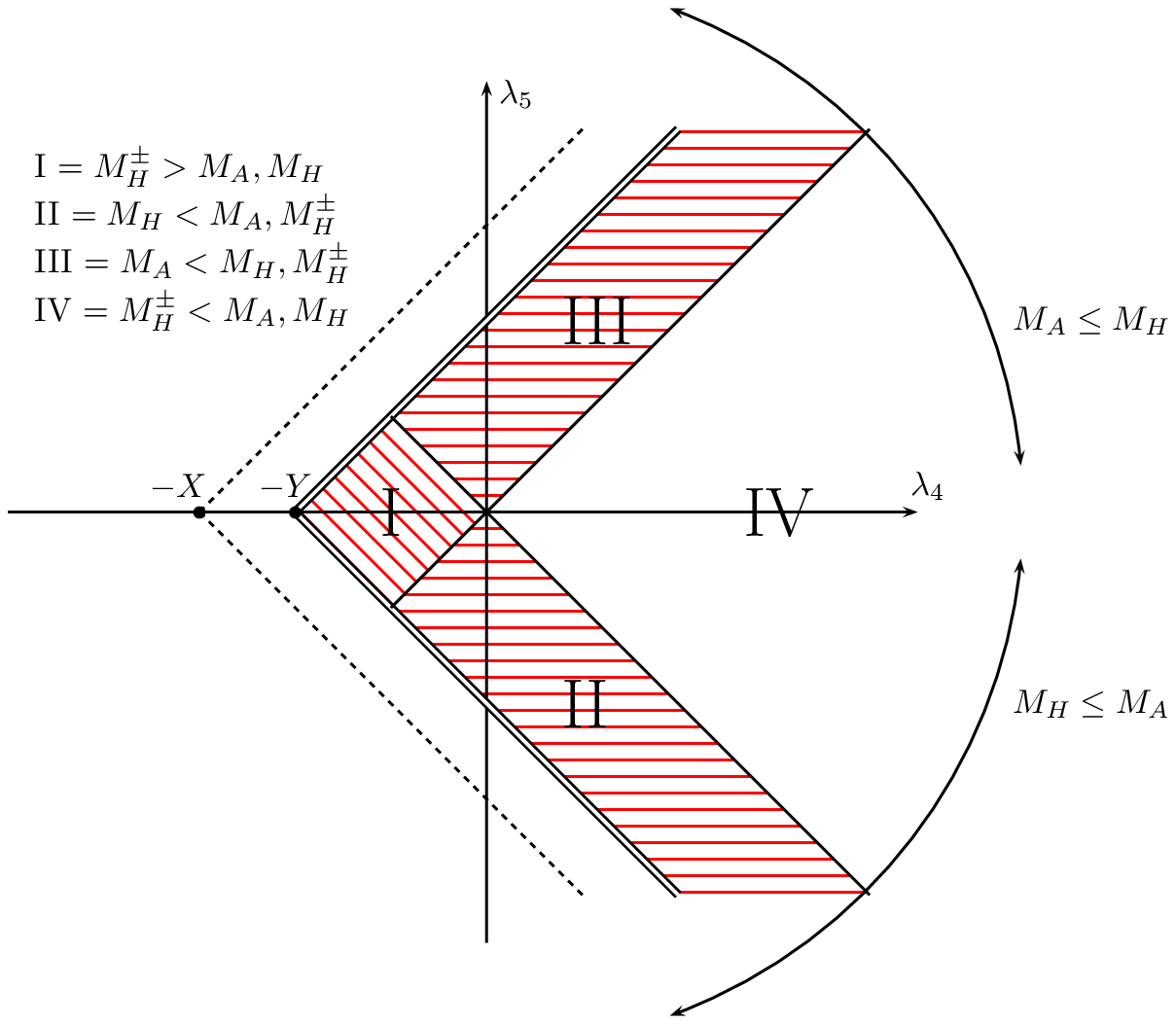,height=7.in}
\vskip -10cm
\caption{The  ($\lambda_4, \lambda_5$) plane. Left:  The N, Ch and Inert sectors. Right: Mass regions for the Inert Model.
In the region I (IV) $H^+$ is heavier (lighter) than both $H$ and $A$ scalars.
The positivity constraints - dotted lines.
\label{fig:dia}}
\end{figure}

The most general vevs in 2HDM with explicit $Z_2$ symmetry are:
 $$
\langle\phi_1\rangle =\left(\begin{array}{c} 0\\
\fr{1}{\sqrt{2}}v_1\end{array}\right), \;\; \langle\phi_2\rangle
=\fr{1}{\sqrt{2}} \left(\begin{array}{c}{u} \\ v_2
 \end{array}\right),\label{genvac}$$
with $v_1,v_2,u$ real, $v^2=v_1^2+{v}_2^2=(246 \,\, \rm{GeV})^2$, $v_1\ge 0$ .
  $Z_2$ is spontaneously broken if  $v_2$ or  $u$ $\neq 0$.
There are three types of solutions: two corresponding to the
$u=0$,   $v_1,v_2 \neq 0$ (Normal extremum), $v_1 \neq 0, v_2=0$ (Inert Model extremum)
and one with $u\neq 0, v_1\neq 0, v_2=0$ (Charge Breaking extremum, with a heavy photon, charge nonconservation, etc.) \cite{fer,kan,mk}.
 A phase diagram in the  $\lambda_4-\lambda_5$ plane shows clearly the corresponding regions  \cite{lorenzo}, see Fig. \ref{fig:dia} (Left).
\section{The Inert Doublet Model}
   In the IDM  $\phi_1$ { is a standard Higgs doublet} and  contains  one physical Higgs boson $h$ with the tree-level couplings to gauge bosons and fermions as in SM. Its  mass is equal to
$$M_h^2=m_{11}^2=\lambda_1 v^2.$$
 {The dark doublet $\phi_2$} contains
four physical spin-0, $Z_2$-odd  particles  $H^\pm, H, A$, called   {dark scalars (collectively denoted by D)}. Their masses (see Fig. \ref{fig:dia}, Right) are given by
$$
M_{H+}^2=-\frac{m_{22}^2}{2}+\frac{\lambda_3}{2}v^2,\,\,\,\,
M_{H}^2=-\frac{m_{22}^2}{2}+\frac{\lambda_3+\lambda_4+\lambda_5}{2}v^2,\,\,\,\,
M_{A}^2=-\frac{m_{22}^2}{2}+\frac{\lambda_3+\lambda_4-\lambda_5}{2}v^2.
$$
 Only quartic couplings which involve solely D depend on (are proportional to) $\lambda_2$.
Quartic and cubic couplings between Higgs boson $h$ and  D's  are
proportional to $M_D^2+m_{22}^2/2$,
those involving $H^\pm$ are proportional to $\lambda_3$ solely.
Dark scalars do couple to W/Z
eg. $H^\pm W^\mp H,\,\,\, H^\pm W^\mp A,\,\,\, AZH.$
\section{Constraints on the Inert Doublet Model}
In all considered versions of 2HDM (beside the 2HDM with charge breaking vacuum)
 there are two charged and three neutral
  physical scalar (spin-0) particles ($h,H$ are CP-even, while $A$ is CP-odd).
  How one can discriminate between various versions of the explicitly $Z_2$-symmetric 2HDM and how existing data can be used to constrain them, especially  IDM?

In many models SM-like scenarios are realized, ie. there exists a light scalar  with mass $>$ 114 GeV  with the SM tree-level couplings and  other non-SM
particles are hard to be observed due to large masses or small couplings to the SM particles.
In the non-supersymmetric 2HDM
both $h$ or $H$ (with the convention that the $h$ is lighter than $H$) can be SM-like.
In the IDM $h$ plays a role of $H_{SM}$ and all the basic tree-level relative couplings $\chi_{u,d,V}=1$.
\paragraph{LEP data}
Recently a dedicated  EW precision test (for oblique parameters S, T,U) for IDM has been
 performed \cite{bar} for $M_h=400-600$ GeV
 with the result:
 $(M_{H^+}-M_A)(M_{H^+}-M_H)=M^2, \,\,\, \, M=120^{+20}_{-30} {\rm {GeV}}.$
 Similar result (small mass splitting)  may hold for $M_h=120-200$, see also \cite{su}.
  We see that $H^\pm$ should be the heaviest particles among D's (region I in Fig. \ref{fig:dia} (Right)).

The absence of a signal within searches for supersymmetric
neutralinos at LEP II was used recently to constrain the IDM \cite{Lundstrom:2008ai}.   This  analysis excludes IDM for  $M_H <$ 80 GeV, \,\, $M_A <$  100 GeV and $\Delta(A,H) >$ 8 GeV.
\paragraph{Testing IDM at colliders}
Deviation from the SM decay rates for $h$ may appear in the IDM due to
the additional decay channels, for a relatively light $H$.
Significant modification of  the Br for $h$ with mass 100-150 GeV may appear, due to $h$ decay to  $H H$, for $M_{H}$  around 50 GeV.
The total width of $h$ is predicted to be enhanced  up to factor 3
for mass of $H^+$ equal 170 GeV and $m_{22}^2=-20$ GeV \cite{cao}.
This effect may be  observed at the LHC, as well as at the planned $e^+e^-$ ILC or PLC
during a SM-Higgs searches.
LHC discovery potential for the dark scalars was studied as well; as the best process
   the  $AH$ production was found eg. \cite{cao,bar,su}.
\paragraph{Dark matter from IDM} In the IDM the dark matter particle could be $H$ (or $A$).
Typically it is taken to be $H$.
 A direct annihilation of HH into $\gamma \gamma$ and $Z\gamma$, for
mass of  DM candidate between 40-80 GeV, was studied in \cite{Gustafsson:2007pc}.
Such DM line signal can be searched for with FERMI satelite.
$M_H$ between 40-80 GeV and  masses of $H^+$ =170 GeV, $A$ =50 -70 GeV, for $M_h$=500 GeV
(and also for $M_h$=120 GeV) were considered.
Other DM study within IDM was performed in \cite{LopezHonorez:2006gr},
 for  $M_h$ =120 GeV and   large $M_{H^+}$, close to $M_A$ = 400 - 550 GeV. Recent work \cite{su} shows that there are  five distinctive regions with a right relic density in the Universe.
 \paragraph{Evolution of the Universe}
 A new type of constraints are coming from the analysis of the possible sequences of phase transitions between various possible vacua after EW symmetry breaking in the early Universe \cite{ii,gkks}. In
  \cite{gkks} the sequence leading to the present Inert phase is analyzed.
\section{Summary}
Among all 2HDMs only the IDM, with SM-like Higgs $h$ and dark scalars $D (H^\pm,\,\,A,\,\,H)$, offers a DM candidate ($H$). The model is in agreement with existing collider and astrophysics data.
\vskip -0.9cm
\vskip +1.1cm
We thank K. Kanishev and I. Ginzburg for a nice collaboration. Work supported by part by
 the EC 6th Framework Programme
MRTN-CT-2006- 035863 and Polish Ministry of Science and Higher Education N N202 230337.

\end{document}